\documentclass[aps,pre,amsmath,amssymb,twocolumn,groupedaddress,showpacs]{revtex4}
\usepackage{graphicx}
\usepackage[utf8]{inputenc}

\usepackage{epsf}

\newcommand{\be}{\begin{equation}}
\newcommand{\ee}{\end{equation}}
\newcommand{\ba}{\begin{eqnarray}}
\newcommand{\ea}{\end{eqnarray}}
\newcommand{\baa}{\begin{eqnarray}}
\newcommand{\eaa}{\end{eqnarray}}
\newcommand{\ed}{\end{document}}

\renewcommand{\baselinestretch}{1.2}
\date{\today}

\begin{document}%large

\title{Dynamics of inertial vortices  in multi-component Bose-Einstein condensates}
\author{Katsuhiro Nakamura$^{(1,2)}$, Doniyor Babajanov$^{(1)}$, Davron Matrasulov$^{(3)}$, Michikazu Kobayashi$^{(4)}$}
\affiliation{$^{(1)}$Faculty of Physics, National University of Uzbekistan, Vuzgorodok, Tashkent 100174,Uzbekistan\\
$^{(2)}$Department of Applied Physics, Osaka City University, Osaka 558-8585, Japan\\
$^{(3)}$Turin Polytechnic University in Tashkent, 17 Niyazov Str., Tashkent 100093, Uzbekistan\\
$^{(4)}$Department of Basic Science, University of Tokyo, Tokyo
153-8902, Japan}

\begin{abstract}
With use of the nonlinear Schr{\"o}dinger (or Gross-Pitaevskii)
equation with strong repulsive cubic nonlinearity, dynamics of
multi-component Bose-Einstein condensates (BECs) with a harmonic trap in 2
dimensions is investigated beyond the Thomas-Fermi regime. In the case when each component has a single vortex, we obtain an effective nonlinear dynamics for vortex cores (particles).
The particles here acquire the inertia, in marked contrast
to the standard theory of point vortices widely known in the usual hydrodynamics.  
The effective dynamics is equivalent to that of charged particles
under a strong spring force and in the presence of Lorentz force with the uniform magnetic field.
The inter-particle (vortex-vortex) interaction is singularly-repulsive and short-ranged with its magnitude decreasing with increasing distance of the center of mass from the trapping center.
"Chaos in the three-body problem" in the three vortices system can be seen, which is not expected in the corresponding point vortices without inertia in 2 dimensions.
\end{abstract}
 \maketitle

\section{Introduction}\label{sec-introduction}

Recently, there has been much interest in theoretical and
experimental studies on trapped atomic Bose-Einstein condensates
(BECs)\cite{rf:peth,rf:pita,rf:kevr}. The
superfluid property of atomic BECs arises from a dual aspect of
waves and particles, i.e., matter waves, and is theoretically
described by the macroscopic wave function. Because of the
nonlinearity of the system caused by interaction between particles,
the macroscopic wave function can take a form of various solitons
such as bright, dark, grey, and vortex solitons, and these solitons
are experimentally observed in BECs.

As for bright solitons, 
Martin {\it et al.}\cite{rf:Mart} theoretically predicted that
the particle-like behavior of three bright solitons in a
one-dimensional $^{87}$Rb BEC was non-integrable and showed its change from
regular motions to chaos. P\'erez-Garc\'{\i}a {\it et
al.}\cite{rf:Monte} applied a variational method to dynamics of bright solitons in a
two-dimensional (2-d) BEC and showed that the center of mass of each soliton obeys 
Newtonian dynamics and Ehrenfest's theorem is
valid if the phase of BEC wave function will be suitably chosen.

As more interesting systems, we can consider solitons in
multi-component BECs which consist of different kinds of atoms or
same kinds of atoms having different spin and have been
experimentally realized. In multi-component BECs, there are not only
intra-component particle interaction but also inter-component
particle interaction which is another origin of nonlinearity, so we
expect novel soliton dynamics which is not seen in single-component
BECs. Yamasaki {\it et al.}\cite{rf:Yamas} developed a variational method to describe bright soliton
dynamics in 2-d multi-components BECs, and
proposed a model of conservative chaos. In 2-d and 3-d systems, however, bright soliton are unstable unless
intra-component interaction oscillates between attraction and
repulsion or coexisting intra-component quintic (three-body) interaction is strong
enough.

On the other hand, topological vortices known as
quantized vortices, i.e.,  topological defects of the
macroscopic wave function can be stable in two dimensions.
Vortices in single-component \cite{rf:nee,rf:frei,midd} and  multi-component \cite{rf:tha,rf:pap,rf:tojo} BECs have been
realized experimentally, giving a good candidate to study
the dynamics of vortices in 2-d and 3-d  BECs.
Most of the theoretical studies, however, are limited to the Thomas-Fermi regime (TFR) in a single-component BEC \cite{fett1,fett2,midd,torr}.
While dynamics of the macroscopic wave function of BEC is described by Gross-Pitaevskii equation (GPE) in Eq.(\ref{eq-GP}) below,
TFR suppresses the kinetic energy part in GPE in the lowest approximation
and therefore the lowest-order  wave function cannot have a healing length which is a hallmark of the vortex core.
It is highly desirable to construct the effective theory of vortices  beyond the TFR.

In this paper, we consider the vortices in 2-d
multi-component BECs beyond the Thomas-Fermi regime. To consider the effective dynamics of
point-like vortices, we extract some degrees of freedom
by using the variational approach, and derive an effective dynamics
with finite degrees of freedom. For the case of a single-component
BEC without a trap, it is well known that the effective hamiltonian for many vortex systems has a standard form
\cite{rf:Neu,rf:Afta,rf:saito}
\begin{equation}
H\sim \sum_{j>i}n_in_j\ln
r_{ij},\label{eq-single-vortex-hamiltonian}
\end{equation}
in the limit of infinitesimal vortex cores. Here $n_i$ is the winding number of the $i$-th vortex, and
$r_{ij}=\sqrt{(x_i-x_j)^2+(y_i-y_j)^2}$ is the distance between
cores of $i$-th and $j$-th vortices.
Equation~(\ref{eq-single-vortex-hamiltonian}) shows that there is no
momentum degree of freedom, and coordinates $x_i$ and $y_i$ are  formally conjugate
each other.
On the other hand, in the case of
multi-component BECs in a trap with each component having a single vortex, we shall see vortices 
acquire momentum degrees of freedom or inertia. 
A very preliminary idea of
the present work was reported in the conference proceedings
\cite{qmc2008}.

This paper is organized as follows. In Sec.~\ref{sec-model},
starting from the multi-component GPE, we apply the variational method with use of
vortex solitons in the Pad\'e approximation, and derive the effective Hamiltonian for vortices.
Confining to the case of two and three vortices, we numerically
investigate the detailed dynamics of vortices in
Sec.~\ref{sec-some-vortices}. There we see a 
chaotic behavior of  the system consisting of three vortices. Section~\ref{sec-conclusion} is
devoted to conclusions and discussions.

\section{Effective nonlinear dynamics generated by the multi-component GPE}\label{sec-model}

In this Section, we consider the trapped multi-component GPE with
vortices and extract some degrees of freedom of vortex soliton by
using a variational technique. 

BEC at zero temperature is described by the GPE. We shall consider a
2-d system of trapped $n$-component macroscopic wave
function $\Phi_1(t,x,y),\Phi_2(t,x,y),\cdots,\Phi_n(t,x,y)$
satisfying the equations
\begin{eqnarray}
 i\frac{\partial}{\partial t}\Phi_i(t,x,y)
&=& \left[ -\nabla^2+V(x,y)+g_{ii}|\Phi_i(t,x,y)|^2\right.\nonumber\\
&+&\left.\sum_{j\ne i}g_{ij}|\Phi_j(t,x,y)|^2\right]\Phi_i(t,x,y),
\label{eq-GP}
\end{eqnarray}
for $i,j=1,\ldots,n$. Here the normalization condition for each component of wave functions is defined by
$\int |\Phi_i(t,x,y)|^2 dxdy=1$ after a proper rescaling of $\Phi_i$ by the particle number $N$ common to all components. 
The effect of trapping is expressed by
$V(x,y)=(x^2+y^2)$. 
Equation (\ref{eq-GP}) is expressed  with use of scaled variables:
using the confining length $l=\sqrt{\frac{\hbar}{m\omega}}$ and
oscillation period $\tau=\omega^{-1}$, space coordinates are scaled
by $l$, time by $2\tau$, wave function by $\frac{1}{l}$, and
nonlinearity by $\frac{\hbar^2}{2m}$. 
The nonlinearity coefficients,
$g_{ij}\equiv 8\pi Na_{ij}/l$ with $a_{ij}$ the scattering length for binary collisions,
are assumed to be positive and much larger than unity.
$g_{ii}$ and $g_{ij}$ with $i\ne j$ stand for intra-component and inter-component interactions, respectively.
We shall choose $g_{ii}=g_1 (\gg 1)$ for all $i$
and $g_{ij}=g_2 (\gg 1)$ for $i\neq j$.

In the absence of the inter-component interaction, each component
has stationary states of a vortex. So, we consider the case in which
each component has one vortex and vortices interact with each other
through the inter-component interaction. Our goal is to
derive from (\ref{eq-GP}) the evolution equation for the collective
coordinates of trial vortex functions (TVFs). The collective coordinates for a
vortex are phase variables besides the coordinates of a vortex core.
We shall use TVF beyond the Thomas-Fermi regime, by incorporating the effect of a kinetic energy in GPE in Eq.(\ref{eq-GP}):
As for the amplitude of TVF, we choose a vortex function based on the Pad\'e approximation \cite{rf:Igor,natali} which is regularized due to a trap. 
As for its phase,
we Taylor-expand the phase with respect to space coordinates around the vortex core. Then TVF
with winding number $n_i=\pm1$ is given by
 
\begin{eqnarray}
&&\Phi_i(t,x,y) \equiv  f_i(t,x,y)\exp[i\phi_i (t,x,y)]\nonumber\\
&&= N\exp \left[-\frac{x^2+y^2}{2\Delta}\right]\sqrt{\frac{(x-x_i)^2+(y-y_i)^2}{2\xi^2+(x-x_i)^2+(y - y_i)^2}}\nonumber\\
&&\times \exp
\left[i\left[n_i\tan^{-1}\left(\frac{y-y_i}{x-x_i}\right)+\alpha_i(x-x_i)+\beta_i(y-y_i)\right]\right]\nonumber\\
\label{eq-trial-function}
\end{eqnarray}
with the normalization factor $N=\frac{1}{\sqrt{\pi\Delta-2\pi
e^{\frac{2\xi^2}{\Delta}}\xi^2\Gamma\left(0,\frac{2\xi^2}{\Delta}\right)}}
$. 
Here $\Gamma(0,z)\equiv\int\limits_z^\infty t^{-1} e^{-t}dt$ is the incomplete gamma function
of the second kind, whose expansion with respect to $z$ is given in Eq.(\ref{in-gamma}) in Appendix \ref{typ-int}.
\newline

The collective coordinates are locations of the core $(x_i,y_i)$ and the first-order
coefficients $(\alpha_i,\beta_i)$ of Taylor-expansion of the phase $\phi_i(t,x,y)$ with
respect to $(x-x_i,y-y_i)$. $\Delta$ in the Gaussian amplitude factor reflects a trap in Eq.(\ref{eq-trial-function}). $\xi$
is the healing length related to vortex core size. The condition
to minimize the energy $E=\int dx dy
\left(|\nabla\Phi_i|^2+V(x,y)|\Phi_i|^2+\frac{g_1}{2}|\Phi_i|^4\right)$
for the individual static component in Eq.(\ref{eq-trial-function}) 
centered at the origin, leads to $\Delta \cong \sqrt{\frac{3}{2}-(\gamma+1)n_i^2+ \frac{g_1}{4\pi}}$ and
$\xi\cong \frac{|n_i|\pi^{1/4}}{\sqrt{2+\gamma}}g_1^{-1/4}
$, respectively, where $\gamma(=0.57721)$ is Euler constant.  We shall use these values for $\Delta$ and $\xi$ in this paper.

The form in Eq.(\ref{eq-trial-function}), which is a product of the vortex solution in the absence of a harmonic trap and Gaussian factor due to the trap, gives a suitable TVF for a vortex under the strong nonlinearity. A different form with use of eigenstates (with non-zero angular momenta) under the 2-d harmonic trap \cite{garc1,garc2,garc3} has no small healing length, results in the inter-vortices force growing with inter-vortices distance, etc, and cannot be suitable as TVF under the strong nonlinearity.

First of all we note: GPE in Eq.(\ref{eq-GP}) can be derived from the
variational principle that minimizes the action obtained from
Lagrangian density $\mathcal{L}$ for field variables,
\begin{align}
-\mathcal{L}=&\sum_i\left[\frac{i}{2}(\Phi_i\dot{\Phi}_i^\ast-\Phi_i^\ast\dot{\Phi}_i)+|\nabla\Phi_i|^2 \right. \nonumber\\
&\left.
+(x^2+y^2)|\Phi_i|^2+\frac{g_1}{2}|\Phi_i|^4\right]+\sum_{j>i}g_2|\Phi_i|^2|\Phi_j|^2.
\label{eq-lagrangian-density}
\end{align}
In fact, Eq.(\ref{eq-GP}) is obtained from Lagrange equation:
\begin{equation}
\frac{\partial}{\partial
t}\frac{\partial\mathcal{L}}{\partial{\dot{\Phi}_i^\ast}}-\frac{\partial\mathcal{L}}{\partial\Phi_i^\ast}+\nabla\frac{\partial\mathcal{L}}{\partial\nabla\Phi_i^\ast}=0.\label{eq-lagrange-equation}
\end{equation}

We now insert TVF in Eq. (\ref{eq-trial-function}) into  Eq.
(\ref{eq-lagrangian-density}).
Noting $(x_i,y_i)$ and $(\alpha_i,\beta_i)$ as time-dependent variables, Eq. (\ref{eq-lagrangian-density}) becomes
\begin{eqnarray}
-\mathcal{L}&=&\sum_i\Bigg[\Bigg(\dot{x}_i\frac{\partial\phi_i}{\partial x_i}+\dot{y}_i\frac{\partial\phi_i}{\partial y_i}+\dot{\alpha}_i\frac{\partial\phi_i}{\partial\alpha_i}+\dot{\beta}_i\frac{\partial\phi_i}{\partial\beta_i}\Bigg)f_i^2 \nonumber\\
&+&\Bigg(\frac{\partial f_i}{\partial
x}\Bigg)^2+\Bigg(\frac{\partial f_i}{\partial y}\Bigg)^2
+\Bigg(\Bigg(\frac{\partial\phi_i}{\partial x}\Bigg)^2+\Bigg(\frac{\partial\phi_i}{\partial y}\Bigg)^2\Bigg)f_i^2\nonumber\\
&+&\frac{g_1}{2}f_i^4+(x^2+y^2)f_i^2\Bigg]+g_2\sum_{j>i}f_i^2f_j^2.
\label{eq-lagrangian-density-variational}
\end{eqnarray}

By integrating $\mathcal{L}$
over space coordinates $(x, y)$, we obtain the effective Lagrangian
$L$ for the collective coordinates:
\begin{equation}
L=\int\int d x d y\mathcal{L}.
\end{equation}
\vspace*{0.5\baselineskip}
In the limit of $\xi^2/\Delta\ll 1$, $L$ is expressed by
\begin{eqnarray}
-&L&=\left(\frac{N}{N_0}\right)^2\sum\limits_{i}\Bigg[-\frac{n_i}{\Delta}\left(\dot{x_i}y_i-\dot{y_i}x_i\right)e^{-\frac{l_{i}^{2}}{\Delta}}\left(d_1+\frac{d_2l_{i}^{2}}{\Delta}\right)\nonumber\\
&+&\left(\alpha_{i}^{2}+\beta_{i}^{2}-\left(\alpha_i\dot{x_i}+\beta_i\dot{y_i}\right)\right)d_1 \nonumber\\
&- &\left(\dot{\alpha_i}x_i+\dot{\beta_i}y_i\right)\left(1-\frac{2\xi^2}{\Delta}\left(1-\frac{l_i^2}{2\Delta}\right)\right)\nonumber\\
&+& \frac{1}{\Delta}e^{-\frac{l_{i}^{2}}{\Delta}}\left(2n_{i}^{2}d_3+d_4+2n_{i}^{2}\frac{l_{i}^{2}d_{5}}{\Delta}+d_6\frac{l_{i}^{2}}{\Delta}\right)\nonumber\\
&+&2\frac{n_i}{\Delta}\left(y_i\alpha_i-x_i\beta_i\right)e^{-\frac{l_{i}^{2}}{\Delta}}\left(d_{1}+\frac{l_{i}^{2}}{\Delta}d_{2}\right) \nonumber\\ 
&+& e^{-\frac{l_{i}^{2}}{\Delta}}\left(d_{8}+l_{i}^{2}d_{9}\right)\nonumber\\
&+&\left(\frac{N}{N_0}\right)^2\frac{g_1}{\pi\Delta}\left(\frac{1}{4}+d_{7}e^{-\frac{2l_{i}^{2}}{\Delta}}\right)
\Bigg] \nonumber\\
&+&\sum\limits_{j>i}U(r_{ij},l_G^{ij}). \label{eq-lagrangian-final}
\end{eqnarray}

Here, $l_i=\sqrt{x_i^2+y_i^2}$ and $N_0=\frac{1}{\sqrt{\pi\Delta}}$. Expressions (:
Taylor-expansions with respect to $\xi^2/\Delta$) for coefficients
$d_1\sim d_{9}$ are listed in Table 1 and the derivation of typical
coefficients is described in Appendix \ref{typ-int}. The expression
for the inter-vortices interaction $U(r_{ij},l_G^{ij})$ , which is a function of inter-vortices distance
$r_{ij}=\sqrt{x_{ij}^2+y_{ij}^2}$ and distance of the center of masses from the origin $\l_G^{ij}=\sqrt{(x_G^{ij})^2+(y_G^{ij})^2}$, will be  given in
Appendix \ref{vort-int}.

Lagrange equations of motion for the phase variables $\alpha_i$ and
$\beta_i$,
\begin{eqnarray}
\frac{d}{dt}\Bigg(\frac{\partial L}{\partial{\dot\alpha_i}}\Bigg)-\frac{\partial L}{\partial\alpha_i}=0\nonumber\\
\frac{d}{dt}\Bigg(\frac{\partial
L}{\partial{\dot\beta_i}}\Bigg)-\frac{\partial L}{\partial\beta_i}=0
\end{eqnarray}
lead to
\begin{eqnarray}
\alpha_i \simeq B_1\dot{x_i}-n_i B_2 y_i\nonumber\\
\beta_i \simeq B_1\dot{y_i}+n_i B_2 x_i.\label{eq-momentum}
\end{eqnarray}

Equation (\ref{eq-momentum}) shows that $(\alpha_i,\beta_i)$
correspond to generalized momentum conjugate to $(x_i,y_i)$ under
the vector potential. Then the equation of motion for $(x_i,y_i)$
\begin{eqnarray}\label{eq-position}
\frac{d}{dt}\Bigg(\frac{\partial L}{\partial{\dot{x}_i}}\Bigg)-\frac{\partial L}{\partial x_i}=0\\
\frac{d}{dt}\Bigg(\frac{\partial
L}{\partial{\dot{y}_i}}\Bigg)-\frac{\partial L}{\partial y_i}=0,
\end{eqnarray}
combined with Eq.(\ref{eq-momentum}), gives
\begin{eqnarray}\label{lorent}
&\ddot{x_i}\simeq n_i B_3 \dot{y_i}-B_4x_i-B_5\sum\limits_{j\neq i}\frac{\partial U(x_{ij},y_{ij})}{\partial x_i}\nonumber\\
&\ddot{y_i}\simeq -n_i B_3 \dot{x_i}-B_4y_i-B_5\sum\limits_{j\neq
i}\frac{\partial U(x_{ij},y_{ij})}{\partial y_i}.
\end{eqnarray}

\begin{widetext}\label{tab1}
{\bf Table 1.} Expressions for coefficients in
Eq.(\ref{eq-lagrangian-final})  (: Taylor-expansions with respect to
$\xi^2/\Delta$) .

\begin{center}
\begin{tabular}{|c|c|} \hline
Coefficients & Expressions  \\ \hline $d_1$&\
$1+\frac{2\gamma\xi^2}{\Delta}+\frac{2\xi^2}{\Delta}\ln{\frac{2\xi^2}{\Delta}}$\\
\hline $d_2$&\ $
\frac{1}{2}-\frac{\xi^2}{\Delta}$\\

\hline $d_3$&\
$-\frac{\gamma}{2}-\frac{1}{2}\ln{\frac{2\xi^2}{\Delta}}+\frac{\xi^2}{\Delta}-\frac{\gamma\xi^2}{\Delta}-\frac{\xi^2}{\Delta}\ln{\frac{2\xi^2}{\Delta}}$\\
\hline $d_4$&\ $\frac{3}{2}+\frac{\xi^2}{\Delta}+\frac{4\gamma\xi^2}{\Delta}+\frac{4\xi^2}{\Delta}\ln{\frac{2\xi^2}{\Delta}}$\\
\hline $d_{5}$&\
$\frac{1}{2}+\frac{\gamma\xi^2}{\Delta}+\frac{\xi^2}{\Delta}\ln{\frac{2\xi^2}{\Delta}}$\\
\hline $d_6$&\
$1+\frac{5\xi^2}{\Delta}+\frac{6\gamma\xi^2}{\Delta}+\frac{6\xi^2}{\Delta}\ln{\frac{2\xi^2}{\Delta}} $\\

\hline $d_{7}$&\
$\frac{\xi^2}{\Delta}+\frac{2\gamma\xi^2}{\Delta}+\frac{2\xi^2}{\Delta}\ln{\frac{4\xi^2}{\Delta}}$\\
\hline $d_{8}$&\
$\Delta-2\xi^2$\\
\hline $d_{9}$&  $1+\frac{2\xi^2}{\Delta}+\frac{2\gamma\xi^2}{\Delta}+\frac{2\xi^2}{\Delta}\ln{\frac{2\xi^2}{\Delta}}$\\
\hline
\end{tabular}

\end{center}
\end{widetext}

Coefficients $B_1\sim B_5$ are given by
\begin{eqnarray}
B_1&=&\frac{d_1-1+\frac{2\xi^2}{\Delta}}{2d_1}\sim\frac{(\gamma+1)\xi^2}{\Delta}\nonumber\\
B_2&=&\frac{1}{\Delta}\nonumber\\
B_3&=&-\frac{4d_1^2}{\Delta(d_1-1+\frac{2\xi^2}{\Delta})^2}\sim -\frac{\Delta}{(\gamma+1)^2\xi^4}\nonumber\\
B_4&=&\frac{4d_1d_{9}}{(d_1-1+\frac{2\xi^2}{\Delta})^2}\sim\frac{\Delta^2}{(\gamma+1)^2\xi^4}\nonumber\\
B_5&=&\left(\frac{N}{N_0}\right)^2\frac{2d_1}{(d_1-1+\frac{2\xi^2}{\Delta})^2}\sim\frac{\Delta^2}{2(\gamma+1)^2\xi^4}.
\label{eq-coefficient-B1-B4}
\end{eqnarray}

Equations (\ref{eq-momentum}) and (\ref{lorent}) are valid aside from a multiplicative global factor $(1 + O(\frac{l_i^2}{\Delta}))$.
The smallness of  $\frac{l_i^2}{\Delta}$ will be justified {\it a posteriori}.
Equation~(\ref{lorent}) shows that dynamics of coordinates
$(x_i,y_i)$ is very similar to charged particles with charges
$n_i=\pm1$ under a strong spring force with a force constant $B_4=O(g_1^2)$ and in the presence of Lorentz force with the magnetic field $\textbf{B}=(0,0,B_3)$. One should note that the spring force here has nothing to do with that of the original harmonic potential $V(x,y)$ in Eq.(\ref{eq-GP}).
The Hamiltonian corresponding to
Eqs.~(\ref{lorent})  can be given by
\begin{equation}
H=\sum_{i}\Big[\frac{1}{2m}[\textbf{p}_i-n_i\textbf{A}_i]^2+W_i+\sum_{j>i}\tilde{U}(r_{ij},l_G^{ij})\Big],\label{eq-hamiltonian}
\end{equation}
with use of the momentum $\textbf{p}_i\equiv m\dot{\textbf{r}} +
n_i\textbf{A}_i$,  unit mass $m=1$, the vector potential
$\textbf{A}_i=\frac{B_3}{2}(-y_i, x_i)$, and the scalar potential
$W_i=\frac{1}{2}B_4(x_i^2+y_i^2)$. 

All values $B_1\sim B_5$ depend on the strength of interaction $g_1$ which is tunable by Feshbach
resonance. In the effective particle dynamics described by Eqs.(\ref{lorent}) and (\ref{eq-hamiltonian}),
the spring constant $B_4$ is large enough to guarantee particles with unit mass to be confined in the neighborhood of the origin, i.e., the trapping center.
This finding  justifies $\frac{l_i^2}{\Delta}\ll 1$ and will also be utilized to obtain the inter-vortices interaction in Appendix \ref{vort-int}.

The scaled inter-vortices
interaction $\tilde{U}(r_{ij},l_G^{ij})$ is given with use of Eq.(\ref{U-final}) as
\begin{eqnarray}\label{magni}
\tilde{U}(r_{ij},l_G^{ij})&=&B_5 U_4(r_{ij},l_G^{ij})\nonumber\\
&\cong& \frac{G}{r_{ij}^2}\exp{\left(-\frac{r_{ij}^2+4(\l_G^{ij})^2}{2\Delta}\right)}
\end{eqnarray}
with the coupling constant
\begin{equation}\label{coup-strength}
G=\frac{4g_2}{(\gamma+1)^2\pi^{1/2}}.
\end{equation}
Therefore the inter-vortices interaction is singularly-repulsive and short-ranged with respect to $r_{ij}$ with its magnitude decreasing with increasing $\l_G^{ij}$.

Compared to Eq.~(\ref{eq-single-vortex-hamiltonian}), it is clear
that the system has momentum degrees of freedom, and vortices have
a behavior of particle with inertia rather than that of vortex point without inertia
widely used in the conventional theory of hydrodynamics \cite{rf:lam,rf:Ons,rf:bat,rf:fris,rf:aref}. This is one of the main assertions of the present paper.
The inertia of vortex appears already in the single-component BEC with a trap.

%Among $B_3(<0),B_4(>0)$ and $B_5(>0)$
%appearing in Eq.(\ref{lorent}), the magnitude of $B_4$ is the
%largest. In solving Eq.(\ref{lorent}), therefore, it is convenient
%to use the scaled variable and coefficients as
%$\tau=\frac{t}{\sqrt{|B_3|}}, \hat{B_3}=\frac{B_3}{|B_3|}=-1,
%\hat{B_4}=\frac{B_4}{|B_3|}$, and $\hat{B_5}=\frac{B_5}{|B_3|}$.  In
%the next Section, in numerically solving the canonical equations
%obtained from Eq.(\ref{eq-hamiltonian}), we shall employ the same
%scaling as above.
In closing this Section we should comment: we also attempted to apply the collective coordinate method using a Laguerre-type trial function which is a good candidate for TVF in the case of a weak nonlinearity \cite{garc1,garc2,garc3}. However, such TVF in the case of a strong nonlinearity  proved to result in: 
1) a time-dependent mass for each vortex and 2) the inter-vortex interaction growing with increasing the inter-vortex distance.
Hence this TVF was not suitable to describe a dynamics of vortices in BECs with a strong nonlinearity.
On the other hand, the effective vortex dynamics in the Thomas-Fermi regime \cite{fett1,fett2,midd,torr}, which suppresses the kinetic energy in constructing TVF, leads to neither non-zero inertia nor Lorentz force.

\section{Dynamics of two and three vortices with inertia}\label{sec-some-vortices}

We shall now focus on the system of two vortices with equal
winding numbers, and see how the trajectory generated by effective particle dynamics 
in Eq.(\ref{lorent}) and (\ref{eq-hamiltonian})
well mimics the orbit of the singular points of wave vortices calculated by using GPE in Eq.(\ref{eq-GP}).
We shall then move to the system of three vortices with equal
winding numbers, and find that chaos appears even in the three vortex
system. This feature is  different from that of point
vortices system in a single component BEC in which chaos can appear
in the case of more than three vortices. 

\subsection{Dynamics of two vortices}

\begin{figure}[htb]
\centering
\includegraphics[width=1.0\linewidth]{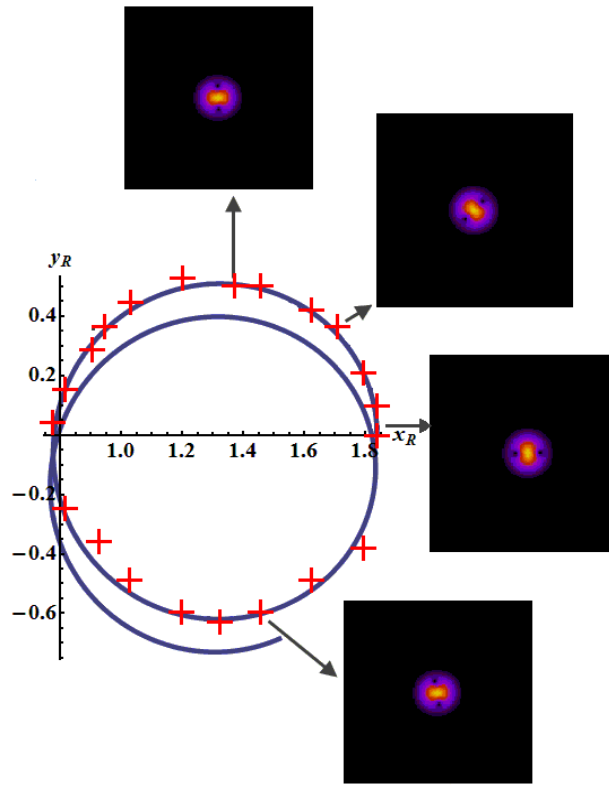}
\caption{Two vortices dynamics with identical winding numbers
$(n_1=n_2=1)$. $g_1=g_2=100$. Initial values: $x_1(0)=y_1(0)=-x_2(0)=-y_2(0)=\frac{1}{\sqrt{2}}$; $\dot{x_1}(0)=\dot{x_2}(0)=0$;  
$\dot{y_1}(0)=-\dot{y_2}(0)=1.1$.
Solid line is a trajectory for relative coordinates of a pair of particles calculated by iterating the canonical equations for the Hamiltonian (\ref{two-hamiltonian-equal}); crosses are corresponding results for a pair of singular points of interacting vortices constructed from the numerical iteration of GPE in Eq.(\ref{eq-GP});  four sub-panels are wave function patterns for interacting vortices.}
\label{fig:traj_color}
\end{figure}

For two vortices with the same winding numbers $n_1=n_2=1$,
Hamiltonian (\ref{eq-hamiltonian}) can be rewritten as
\begin{eqnarray}
H&=&\frac{1}{2}(p_{x,T}^2+p_{y,T}^2+p_{x,R}^2+p_{y,R}^2)\nonumber\\
&-&\frac{B_3}{2}(p_{x,T}y_T-p_{y,T}x_T+p_{x,R}y_R-p_{y,R}x_R)\nonumber\\
&+&\left(\frac{B_3^2}{8}+\frac{B_4}{2}\right)(x_T^2+y_T^2+x_R^2+y_R^2)\nonumber\\
%&+&\tilde{U}\left(\sqrt{2(x_R^2+y_R^2)},\sqrt{(x_T^2+y_T^2)/2}\right)\nonumber\\
&+&\frac{G}{2(x_R^2+y_R^2)}\exp\left(-\frac{x_R^2+y_R^2+x_T^2+y_T^2}{\Delta}\right),
\label{two-hamiltonian-equal}
\end{eqnarray}
where $(x_T,y_T)$ and $(x_R,y_R)$ play the role of the center-of-mass and relative coordinates, respectively and $(p_{x,T},p_{y,T})$ and $(p_{x,R},p_{y,R})$
are their canonical-conjugate variables. To be explicit,
\begin{eqnarray}
(x_T,y_T)&=&\frac{1}{\sqrt{2}}(x_1+x_2,y_1+y_2)\nonumber\\
(x_R,y_R)&=&\frac{1}{\sqrt{2}}(x_1-x_2,y_1-y_2)\nonumber\\
(p_{x,T},p_{y,T})&=&\frac{1}{\sqrt{2}}(p_{x,1}+p_{x,2},p_{y,1}+p_{y,2})\nonumber\\
(p_{x,R},p_{y,R})&=&\frac{1}{\sqrt{2}}(p_{x,1}-p_{x,2},p_{y,1}-p_{y,2}).
\label{two-coordinate}
\end{eqnarray}

Hamiltonian (\ref{eq-hamiltonian}) cannot be reduced to two independent
two degree-of-freedom subsystems (: center-of-mass system and
relative-coordinates system) because the inter-vortex interaction depends not only on $r_{ij}$ but also on $l_G^{ij}$. However, we see mostly KAM tori in this system.
As shown in Fig. \ref{fig:traj_color}, we find that the trajectory of $(x_R, y_R)$ generated by effective two-particle dynamics
well mimics the corresponding orbit obtained by a pair of singular points calculated by using GPE in Eq.(\ref{eq-GP}). This fact justifies the validity of our trial function in Eq.(\ref{eq-trial-function}) and the resultant equation of motion for collective coordinates in Eq.(\ref{lorent}) and (\ref{eq-hamiltonian}).

\subsection{Dynamics of three vortices}

Encouraged by the effectiveness of the collective coordinate method in the case of two vortices, we proceed to the dynamics of three vortices with the identical winding numbers
$n_1=n_2=n_3=1$, whose Hamiltonian (\ref{eq-hamiltonian}) becomes
\begin{eqnarray}\label{three-hamil}
H&=&\frac{1}{2}\left(p_{xC}^2+p_{yC}^2+p_{xT}^2+p_{yT}^2+p_{xR}^2+p_{yR}^2\right)\nonumber\\
&-&\frac{B_3}{2}(x_C p_{yC}+x_R p_{yR}+x_T p_{yT}\nonumber\\
&-&y_C p_{xC}-y_R p_{xR}-y_T p_{xT} ) \nonumber\\
&+& \left(\frac{B_3^2}{8}+\frac{B_4}{2}\right)\left(x_C^2+x_R^2+x_T^2+y_C^2+y_R^2+y_T^2\right)\nonumber\\
&+&\sum_{(i,j)=(1,2),(2,3),(3,1)}\frac{G}{(x_i-x_j)^2+(y_i-y_j)^2}\nonumber\\
&\times&\exp\left(-\frac{x_i^2+y_i^2+x_j^2+y_j^2}{\Delta}\right),
%&+&\tilde{U}\Big(\sqrt{((\sqrt{3}x_R+x_C)^2+(\sqrt{3}y_R+y_C)^2))/2},\nonumber\\
%&&\sqrt{((2\sqrt{2}x_T-x_R+\sqrt{3}x_C)^2+(2\sqrt{2}y_T-y_R+\sqrt{3}y_C)^2)/24}\Big)\nonumber \\
%&+&\tilde{U}\Big(\sqrt{((\sqrt{3}x_R-x_C)^2+(\sqrt{3}y_R-y_C)^2))/2},\nonumber\\
%&&\sqrt{((2\sqrt{2}x_T-x_R-\sqrt{3}x_C)^2+(2\sqrt{2}y_T-y_R-\sqrt{3}y_C)^2)/24}\Big)\nonumber \\
%&+&\tilde{U}\Big(\sqrt{2(x_C^2+y_C^2)},\nonumber\\
%&&\sqrt{((\sqrt{2}x_T+x_R)^2+(\sqrt{2}y_T+y_R)^2)/6}\Big),
%&-&G\exp\left( \frac{3((\sqrt{3}x_R+x_C)^2+(\sqrt{3}y_R+y_C)^2))+(2\sqrt{2}x_T-x_R+\sqrt{3}x_C)^2
%+(2\sqrt{2}y_T-y_R+\sqrt{3}y_C)^2}{12\Delta}\right)
\end{eqnarray}
where Jacobi coordinates ($x_T,y_T,etc$) are defined by
\begin{eqnarray}
(x_T,y_T)&=&\frac{1}{\sqrt{3}}(x_1+x_2+x_3,y_1+y_2+y_3)\nonumber\\
(x_C,y_C)&=&\frac{1}{\sqrt{2}}(x_1-x_3,y_1-y_3),\nonumber\\
(x_R,y_R)&=&\frac{1}{\sqrt{6}}(x_1+x_3-2x_2,y_1+y_3-2y_2)\nonumber\\
(p_{xR},p_{yR})&=&\frac{1}{\sqrt{6}}(p_{x1}+p_{x3}-2p_{x2},p_{y1}+p_{y3}-2p_{y2})\nonumber\\
(p_{xT},p_{yT})&=&\frac{1}{\sqrt{3}}(p_{x1}+p_{x2}+p_{x3},p_{y1}+p_{y2}+p_{y3})\nonumber\\
(p_{xC},p_{yC})&=&\frac{1}{\sqrt{2}}(p_{x2}-p_{x1},p_{y2}-p_{y1}),
\label{three-coordinate}
\end{eqnarray}
which conserve the canonical structure
($\{x_T,p_{xT}\}=\frac{1}{3}\sum\limits_{j=1}^3\{x_j,p_{xj}\}=1,$
etc.). Here $(x_T, y_T)$, $(x_C, y_C)$, and $(x_R,y_R)$  represent
the center of mass of all three components, the relative
displacement, and the bisector of the vertex $(x_2, y_2)$,
respectively. In marked contrast to the massless three vortex system in 2 dimensions which is integrable,  all 6 degrees of freedom are coupled
and the number of independent constants of motion is 2 (energy and $z$-component of angular momentum). Then
Poincar\'e-Bendixon's theorem guaranties the nonintegrability and
chaos of the three inertial vortex system \cite{comme}.

\begin{figure}[htb]
%\centerline{\includegraphics[width=\columnwidth]{poincare.eps}}
%\centering
%\includegraphics[width=1.2\linewidth]{poincare1.eps}
%\includegraphics[width=1.2\linewidth]{poincare2.eps}
%\centerline{\includegraphics[width=1.1\linewidth]{poincare1.eps}}
%\centerline{\includegraphics[width=1.1\linewidth]{poincare2.eps}}
%\centerline{\includegraphics[width=\columnwidth]{poincare1.eps}}
%\centerline{\includegraphics[width=\columnwidth]{poincare2.eps}}
\centerline{\includegraphics[width=\columnwidth]{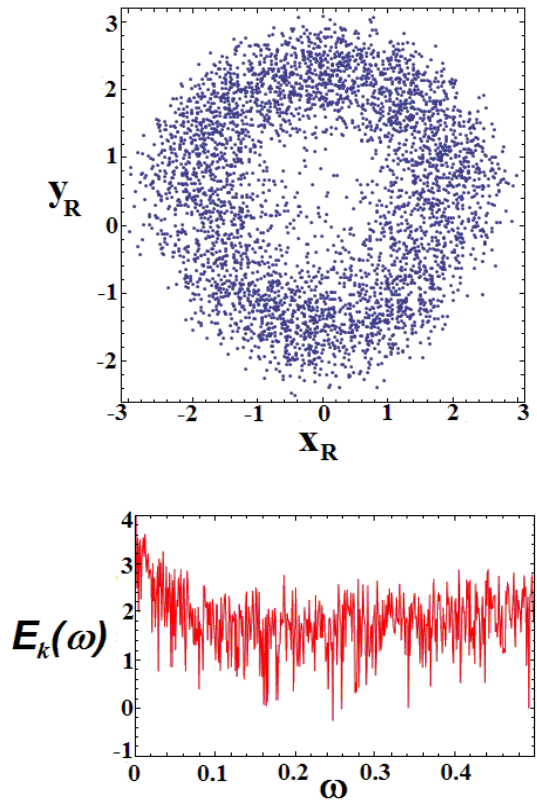}}
\caption{Three vortices dynamics with identical winding numbers
($n_1=n_2=n_3=1$).  $g_1=g_2=100$. Initial values: $x_R(0)=x_C(0)=x_T(0)=1$, $y_R(0)=-1$, $y_C(0)=y_T(0)=1$;
$p_{x,R}(0)=p_{x,C}(0)=p_{x,T}(0)=p_{y,R}(0)=p_{y,C}(0)=p_{y,T}(0)=1$. (a) Poincar\'e cross section; (b)
Power spectra for kinetic energy $E_k(t)$.} \label{fig:poincare3}
\end{figure}

We can construct from
(\ref{three-hamil}) the canonical equations of motion for $x_T, y_T, x_C, y_C, x_R, y_R$
and their canonical-conjugate variables, which is solved
numerically. Poincar\'e cross section
and power spectra for the kinetic energy $E_k(t)$ in
Fig.~\ref{fig:poincare3} give a clear evidence of high-dimensional
chaos. Because of the short-range nature of the interaction, 
we see the emergence of chaos in low-lying energy regions where vortices often meet each other.

\section{Conclusion}\label{sec-conclusion}
We explored vortex dynamics in the 2-dimensional multi-component BEC in the
harmonic trap in the case that each component has a single vortex.
The investigation beyond the Thomas-Fermi regime is made on the nonlinear Schr{\"o}dinger equation with strong repulsive cubic nonlinearity.
With use of a trial vortex function based on the Pad\'e approximation which is regularized due to a trap, we applied a collective coordinates method, obtaining an effective nonlinear dynamics for
vortex cores (particles), which is equivalent to charged particles with inertia under a strong spring force and in the presence of Lorentz force with the uniform magnetic field. The inter-particle interaction is singularly-repulsive and short-ranged with its magnitude decreasing with increasing distance of the center of mass from the trapping center.
The most important finding is the nonzero inertia of vortices, which is not present
in the conventional theory of point vortices widely used in the usual hydrodynamics
\cite{rf:lam,rf:Ons,rf:bat,rf:fris,rf:aref}. The system
of three vortices with inertia can be chaotic, in contrast to the corresponding case of point vortices without inertia.

{\em Acknowledgments.} One of the authors (K. N.) is grateful to F. Abdullaev,  B.
Baizakov, S. Dmitriev and M. Tsubota for useful comments in various important stages of this work.

\appendix
\section{Calculation of typical integrals}\label{typ-int}
We shall calculate some integrals used in this paper.

\subsection{Integral $A\equiv\int\int dx dy f_i^2\left(\dot{x_i}\frac{\partial \phi_i}{\partial x_i}+\dot{y_i}\frac{\partial \phi_i}{\partial y_i}\right)$}
Substituting Eq.(\ref{eq-trial-function}) into the above integrand,
we find
\begin{eqnarray}\label{AAA}
A&=&\frac{1}{\pi\Delta}\left(\frac{N}{N_0}\right)^2\int
\limits_{-\infty}^{\infty}\int \limits_{-\infty}^{\infty}dx dy
e^{-\frac{x^2+y^2}{\Delta}}\nonumber\\
&\times&\Bigg[
n_i\frac{\dot{x_i}(y-y_i)-\dot{y_i}(x-x_i)}{2\xi^2+(x-x_i)^2+(y-y_i)^2} \nonumber\\
&-& (\alpha_i\dot{x_i}+\beta_i\dot{y_i})\frac{(x-x_i)^2+(y-y_i)^2 }{2\xi^2+(x-x_i)^2+(y-y_i)^2}\Bigg],\nonumber\\
\end{eqnarray}
where $N_0=\frac{1}{\sqrt{\pi\Delta}}$ and $N$ is the normalization constant defined below
Eq.(\ref{eq-trial-function}).
Equation(\ref{AAA}) is a sum of $n_i$-dependent term ($A_1$) and $(\alpha_i, \beta_i)$-dependent term ($A_2$). Below we shall concentrate on $A_1$ because $A_2$ is easily calculable. Using polar coordinates as
\begin{eqnarray}
x-x_i=\rho\cos{\theta}&x_i=l_i\cos{\phi_i}\nonumber\\
y-y_i=\rho\sin{\theta}&y_i=l_i\sin{\phi_i}\nonumber\\
dxdy=\rho d\rho d\theta& \label{polar-coordinates}
\end{eqnarray}
we rewrite the integral $A_1$ as
\begin{eqnarray}
A_1&=&\frac{n_i}{\pi\Delta}\left(\frac{N}{N_0}\right)^2\int
\limits_{0}^{\infty}d\rho
e^{-\frac{\rho^2+l_i^2}{\Delta}}\frac{\rho^2}{2\xi^2+\rho^2}\nonumber\\
&\times&\int
\limits_{0}^{2\pi}d\theta(\dot{x_i}\sin{\theta}-\dot{y_i}\cos{\theta})e^{-\frac{2\rho
l_i}{\Delta}\cos{(\theta-\phi_i)}}.
\end{eqnarray}

The $\theta$-integration gives
\begin{eqnarray}
\int
\limits_{0}^{2\pi}d\theta(\dot{x_i}\sin{\theta}-\dot{y_i}\cos{\theta})e^{-\frac{2\rho
l_i}{\Delta}\cos{(\theta-\phi_i)}}\nonumber\\
=\int
\limits_{0}^{2\pi}d\tilde{\theta}\Bigg [\left(\dot{x}\cos{\phi_i}+\dot{y}\sin{\phi_i}\right)\sin{\tilde{\theta}}\nonumber\\
+\left(\dot{x}\sin{\phi_i}-\dot{y}\cos{\phi_i}\right)\cos{\tilde{\theta}}\Bigg]e^{-\frac{2\rho
l_i}{\Delta}\cos{\tilde{\theta}}}\nonumber\\
=-\pi(\dot{x_i}y_i-\dot{y_i}x_i)\left(\frac{2\rho}{\Delta}+\frac{\rho^3l_i^2}{\Delta^3}\right).
\end{eqnarray}
Here we took $\tilde{\theta}=\theta-\phi_i$ and  used the formulas like
\begin{equation}
\int \limits_{0}^{2\pi}d\tilde{\theta} \cos{\tilde{\theta}}e^{-z\cos{\tilde{\theta}}}=-2\pi
I_1(z),
\end{equation}
where $I_1(z)$ is the modified Bessel function, which is expanded as
\begin{equation}
I_1(z)=\frac{z}{2}+\frac{z^3}{16}+O(z^5).
\end{equation}

As for $\rho$-integration, we use  identities like
\begin{equation}
\frac{\rho^3}{2\xi^2+\rho^2}=\rho-\frac{2\xi^2\rho}{2\xi^2+\rho^2}
\end{equation}
and then apply the formulas
\begin{equation}
\int\limits_{0}^\infty d\rho \rho^n
e^{-\frac{\rho^2}{\Delta}}=\frac{1}{2}\Delta^{\frac{n+1}{2}}\Gamma(\frac{n+1}{2})
\end{equation}
and
\begin{eqnarray}
&&\int\limits_{0}^\infty d\rho
e^{-\frac{\rho^2}{\Delta}}\frac{\rho}{2\xi^2+\rho^2}=\frac{1}{2}e^{\frac{2\xi^2}{\Delta}}\Gamma\left(0,\frac{2\xi^2}{\Delta}\right)\nonumber\\
&=&\frac{1}{2}e^{\frac{2\xi^2}{\Delta}}\left(-\gamma-\ln{\frac{2\xi^2}{\Delta}}+\frac{2\xi^2}{\Delta}-\left(\frac{\xi^2}{\Delta}\right)^2\right).
\end{eqnarray}
$\Gamma(0,z)\equiv\int\limits_z^\infty t^{-1} e^{-t}dt$ above is the incomplete gamma function of the second kind and can be expanded as
\begin{equation}\label{in-gamma}
\Gamma(0,z)=-\gamma-\ln{z}+z-\frac{z^2}{4}+\frac{z^3}{18}+O(z^4).
\end{equation}

The final result is
\begin{eqnarray}\label{AAA1}
A_1=-\left(\frac{N}{N_0}\right)^2\frac{n_i}{\Delta}(\dot{x_i}y_i-\dot{y_i}x_i)e^{-\frac{l_i^2}{\Delta}}\left(d_1+\frac{d_2 l_i^2}{\Delta}\right),\nonumber\\
\end{eqnarray}
where $d_1$ and $d_2$ are listed in Table 1.
Equation (\ref{AAA1}), combined with
\begin{eqnarray}\label{AAA2}
A_2&=&-\left(\frac{N}{N_0}\right)^2 (\alpha_i\dot{x_i}+\beta_i\dot{y_i})d_1,
\end{eqnarray}
gives rise to the calculated result for $A$ in Eq.(\ref{AAA}).

\subsection{Integral $B\equiv \frac{g_1}{2}\int\int f_i^4 dx dy $}
\begin{eqnarray}
B&=&\frac{g_1}{2\pi^2\Delta^2}\left(\frac{N}{N_0}\right)^4\int
\limits_{-\infty}^{\infty}\int \limits_{-\infty}^{\infty}dx dy
e^{-\frac{2(x^2+y^2)}{\Delta}}\nonumber\\
&\times&\left(\frac{(x-x_i)^2+(y-y_i)^2}{2\xi^2+(x-x_i)^2+(y-y_i)^2}\right)^2.
\end{eqnarray}
Using polar coordinates in (\ref{polar-coordinates}), 
we rewrite the integral as
\begin{eqnarray}
B&=&\frac{g_1}{2\pi^2\Delta^2}\left(\frac{N}{N_0}\right)^4\int
\limits_{0}^{\infty}\rho d\rho
e^{-\frac{2(\rho^2+l_i^2)}{\Delta}}\nonumber\\
&\times&\left(\frac{\rho^2}{2\xi^2+\rho^2}\right)^2\int
\limits_{0}^{2\pi}d\theta e^{-\frac{4\rho
l_i}{\Delta}\cos{(\theta-\phi_i)}}.
\end{eqnarray}
First, we carry out the $\theta$-integration. Note that
\begin{equation}
\int \limits_{0}^{2\pi}d\theta e^{-z\cos{\theta}}=2\pi I_0(z)
\end{equation}
where $I_0(z)$ is the modified Bessel function, which is expanded as
\begin{equation}
I_0(z)=1+\frac{z^2}{4}+\frac{z^4}{64}+O(z^6).
\end{equation}

Concerning the $\rho$-integration, we first employ
the decomposition
\begin{equation}
\frac{\rho^4}{(2\xi^2+\rho^2)^2}=1-\frac{4\xi^2}{2\xi^2+\rho^2}+\frac{4\xi^4}{(2\xi^2+\rho^2)^2},
\end{equation}
and then apply
the formulas like 
\begin{equation}
\int\limits_{0}^\infty d\rho \rho
e^{-\frac{2\rho^2}{\Delta}}I_0\left(\frac{4\rho
l_i}{\Delta}\right)=\frac{\Delta}{4}e^{\frac{2l_i^2}{\Delta}}.
\label{int-bessel}
\end{equation}

The final result is
\begin{eqnarray}
B&=&\left(\frac{N}{N_0}\right)^4\frac{g_1}{\pi\Delta}
\Bigg[\frac{1}{4}
+d_7e^{-\frac{2l_i^2}{\Delta}}\Bigg], \nonumber\\
\end{eqnarray}
where $d_7$ is given in Table 1.

\section{Calculation of inter-vortices interaction $U$}\label{vort-int}
This interaction is due to the integral
\begin{eqnarray}
U&\equiv&
g_2\int\limits_{-\infty}^{\infty}\int\limits_{-\infty}^{\infty} f_i^2 f_j^2 dx dy\nonumber\\
&=&\frac{g_2}{
\pi^2\Delta^2}\left(\frac{N}{N_0}\right)^4\int\limits_{-\infty}^{\infty}\int\limits_{-\infty}^{\infty}dx
dy e^{-\frac{2(x^2+y^2)}{\Delta}}\nonumber\\
&\times&\frac{(x-x_i)^2+(y-y_i)^2}{2\xi^2+(x-x_i)^2+(y-y_i)^2}\nonumber\\
&\times&\frac{(x-x_j)^2+(y-y_j)^2}{2\xi^2+(x-x_j)^2+(y-y_j)^2}.
\label{integral-17}
\end{eqnarray}
With use of  a prescription $U\equiv \frac{g_2}{\pi^2\Delta^2}\left(\frac{N}{N_0}\right)^4\hat{U}$, the integral becomes:
\begin{eqnarray}
&\hat{U}&=\int\limits_{-\infty}^{\infty}\int\limits_{-\infty}^{\infty}dx
dy e^{-\frac{2(x^2+y^2)}{\Delta}}\nonumber\\
&\times&\Bigg[1-2\xi^2\left(\frac{1}{2\xi^2+(x-x_i)^2+(y-y_i)^2}+(i\rightarrow j)\right)\nonumber\\
%&+&\frac{4\xi^4}{\left(2\xi^2+(x-x_i)^2+(y-y_i)^2\right)\left(2\xi^2+(x-x_j)^2+(y-y_j)^2\right)}\Bigg]\nonumber\\
&+&\frac{4\xi^4}{\left(2\xi^2+(x-x_i)^2+(y-y_i)^2\right)\left(\quad i \rightarrow j \quad \right)}\Bigg]\nonumber\\
&\equiv&\hat{U}_0+\hat{U}_2+\hat{U}_4.
\label{inter-decomp}
\end{eqnarray}
$\hat{U}_0$, $\hat{U}_2$ and $\hat{U}_4$ corresponds to the contributions from $O(\xi^0)$, $O(\xi^2)$ and $O(\xi^4)$, respectively. Among them, $\hat{U}_4$ is responsible to the vortex-vortex interaction, which we shall calculate below.

\begin{figure}[htb]
\centering
\includegraphics[width=1.0\linewidth]{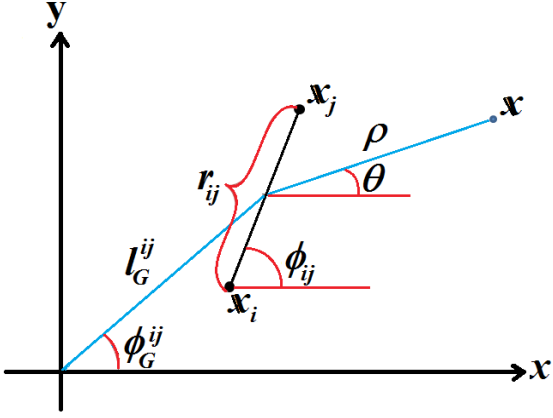}
\caption{New integration variables, center-of-mass and relative coordinates.}
\label{fig:angle}
\end{figure}

Let's define the center-of-mass and relative coordinates by
\begin{equation}\label{com cood}
x_G^{ij}=\frac{x_j+x_i}{2}; \quad y_G^{ij}=\frac{y_j+y_i}{2}
\end{equation}
and 
\begin{equation}
x_i-x_j=r_{ij}\cos{\phi_{ij}}; \quad y_i-y_j=r_{ij}\sin{\phi_{ij}},
\end{equation}
respectively, and transform the integration variables to new ones as (see Fig.\ref{fig:angle})
\begin{eqnarray}
x-x_G^{ij}&=&\rho\cos{\theta}, \quad x_G^{ij}=l_G^{ij}\cos{\phi_G^{ij}}\nonumber\\
y-y_G^{ij}&=&\rho\sin{\theta}, \quad y_G^{ij}=l_G^{ij}\sin{\phi_G^{ij}}\nonumber\\
dxdy&=&\rho d\rho d\theta.
\end{eqnarray}
%\begin{equation}
%x^2+y^2=\rho^2+l_G^2+2\rho l_G\cos(\theta-\phi_G),
%\end{equation}
%\begin{equation}
%(x-x_i)^2+(y-y_i)^2=\rho^2+\frac{r_{ij}^2}{4}+\rho
%r_{ij}\cos(\theta-\phi_{ij}),
%\end{equation}
%\begin{equation}
%(x-x_j)^2+(y-y_j)^2=\rho^2+\frac{r_{ij}^2}{4}-\rho
%r_{ij}\cos(\theta-\phi_{ij}).
%\end{equation}
Then $\hat{U}_4$ in Eq. (\ref{inter-decomp}) becomes
\begin{eqnarray}
%U&=&g_2\pi^2\Delta^2\left(\frac{N}{N_0}\right)^4\int\limits_0^{\infty}\rho
%d\rho\nonumber\\
%&\times&\int\limits_0^{2\pi}
%d\theta e^{\frac{-2}{\Delta}\left(l_G^2+\rho^2+2\rho l_G \cos{(\theta-\phi_G)}\right)}\nonumber\\
%&\times&\frac{\rho^2+\frac{r_{ij}^2}{4}+\rho
%r_{ij}\cos{(\theta-\phi_{ij})}}{2\xi^2+\rho^2+\frac{r_{ij}^2}{4}+\rho
%r_{ij}\cos{(\theta-\phi_{ij})}}\nonumber\\
%&\times&\frac{\rho^2+\frac{r_{ij}^2}{4}-\rho
%r_{ij}\cos{(\theta-\phi_{ij})}}{2\xi^2+\rho^2+\frac{r_{ij}^2}{4}-\rho
%r_{ij}\cos{(\theta-\phi_{ij})}}\nonumber\\
%r_{ij}\cos{(\theta-\phi_{ij})}}\nonumber\\
\hat{U}_4&=&\int\limits_0^{\infty}\rho
d\rho\int\limits_0^{2\pi}
d\theta' e^{\frac{-2}{\Delta}\left((l_G)^2+\rho^2+2\rho l_G^{ij} \cos{(\theta'+\phi_{ij}-\phi_G^{ij})}\right)}\nonumber\\
&\times&\Bigg[\frac{4\xi^4}{(2\xi^2+\rho^2+\frac{r_{ij}^2}{4})^2-\rho^2 r_{ij}^2\cos{^2(\theta')}}\Bigg],
\label{potential-polar}
\end{eqnarray}
where we moved to a new angle variable $\theta'\equiv\theta-\phi_{ij}$.
%We shall first carry out the $\theta$ integration and then $\rho$ integration. The $\theta$ integration of the first %term is
%\begin{equation}
%\int\limits_0^{2\pi}d\theta e^{-\frac{4\rho
%l_G^{ij}}{\Delta}\cos{(\theta-\phi_G^{ij})}}=\int\limits_0^{2\pi}d\tilde{\theta}
%e^{-\frac{4\rho l_G^{ij}}{\Delta}\cos{\tilde{\theta}}}=2\pi
%I_0\left(\frac{4\rho l_G^{ij}}{\Delta}\right).
%\end{equation}
%The integration of the second and third terms in Eq.(\ref{potential-polar}) are identical.
Using the expansion
\begin{equation}
e^{-X \cos{(\theta'+\alpha)}}=\sum\limits_{n=-\infty}^{\infty}(-1)^n I_n(X) e^{i n\alpha}e^{i n\theta'}
\end{equation}
in Eq.(\ref{potential-polar}) and keeping the $n=0$ term,  the integration over the angle variable leads to:
\begin{eqnarray}\label{loren-compl}
&&\int\limits_0^{2\pi}d\theta'\frac{4\xi^4}{(2\xi^2+\rho^2+\frac{r_{ij}^2}{4})^2-\rho^2
r_{ij}^2\frac{1+\cos{2\theta'}}{2}}\nonumber\\
&=&\frac{8\pi\xi^4}{2\xi^2+\rho^2+\frac{r_{ij}^2}{4}}\nonumber\\
&\times&\frac{1}{\sqrt{\left(2\xi^2+\left(\rho-\frac{r_{ij}}{2}\right)^2\right)\left(2\xi^2+\left(\rho+\frac{r_{ij}}{2}\right)^2\right)}},
\end{eqnarray}
where we used the formula 
$\int\limits_0^{2\pi}d\theta\frac{1}{c+b\cos{\theta}}=\frac{2\pi}{\sqrt{(c-b)(c+b)}}$.

We shall proceed to $\rho$ integration. Here  we should note: In the case $\xi\ll1$,  
the Lorentzian-like function on r.h.s. of Eq.(\ref{loren-compl}) is sharply peaked around $\rho=\frac{r_{ij}}{2}$ and  is well approximated for $\rho>0$ by Gaussian, $\frac{8\sqrt{2}\pi\xi^3}{r_{ij}^3}\exp{\left(-\frac{\left(\rho-\frac{r_{ij}}{2}\right)^2}{4\xi^2}\right)}$.  
Therefore,  
the $\rho$ integration leads to:
\begin{eqnarray}
\hat{U}_4&=&\frac{8\sqrt{2}\pi\xi^3}{r_{ij}^3}e^{-\frac{2(l_G)^2}{\Delta}}\nonumber\\
&\times&\Bigg[\int\limits_0^{\infty}\rho
e^{-\frac{2\rho^2}{\Delta}}I_0\left(\frac{4\rho l_G^{ij}}{\Delta}\right)\exp{\left(-\frac{\left(\rho-\frac{r_{ij}}{2}\right)^2}{4\xi^2}\right)}
d\rho\Bigg]\nonumber\\
%&=&\frac{\pi\Delta}{2}-4\xi\frac{\sqrt{2}\pi}{r_{ij}}e^{-\frac{2l_G^2}{\Delta}}\int\limits_0^{\infty}\rho
%e^{-\frac{2\rho^2}{\Delta}}I_0\left(\frac{4\rho l_G}{\Delta}\right)\nonumber\\
%&\times&\exp{\left(-\frac{\left(\rho-\frac{r_{ij}}{2}\right)^2}{4\xi^2}\right)}
%d\rho,\nonumber\\
&=&\frac{8\pi^{3/2}\xi^4}{(1+\xi^2/\Delta)(1+8\xi^2/\Delta)^{1/2}} \frac{1}{r_{ij}^2}
 I_0\left(\frac{2l_G^{ij} r_{ij}}{\Delta (1+2\xi^2/\Delta)}\right)\nonumber\\
& \times&\exp{\left(-\frac{2l_G^2}{\Delta}\right)} \exp{\left(-\frac{r_{ij}^2}{2\Delta(1+8\xi^2/\Delta)}\right)}, 
\end{eqnarray}
where the integration was carried out by the saddle-point approximation which is justified in the case $\xi \ll 1$.
Since each vortex dynamics occurs in the range $\frac{2l_G^{ij} r_{ij}}{\Delta}\ll 1$ because of the notion above Eq.(\ref{magni}) in Sec.\ref{sec-model}
, we may approximate $ I_0\left(\frac{2l_G^{ij} r_{ij}}{\Delta (1+2\xi^2/\Delta)}\right)\sim 1$ and neglect the contributions from $I_n(x)$ with $n=1,2.\cdots$ in the region $x \ll 1$. 
Then we reach
\begin{eqnarray}\label{U-final}
U_4&\equiv&U_4(r_{ij},l_G^{ij})\equiv \frac{g_2}{\pi^2\Delta^2}\left(\frac{N}{N_0}\right)^4\hat{U}_4\nonumber\\
&\approx&\frac{8g_2\xi^4}{\pi^{1/2}\Delta^2}\frac{1}{r_{ij}^2}\exp{\left(-\frac{r_{ij}^2+4(l_G^{ij})^2}{2\Delta}\right)}
\end{eqnarray}
with use of $\left(\frac{N}{N_0}\right)^4\approx 1$.
At first the magnitude of the inter-vortices interaction looks very small (i.e., $O(\xi^4)$), but,
after scaling to make unity the inertial mass of each vortex, it becomes
$O(g_2)$ (see Eq.(\ref{coup-strength})). It is interesting that the inter-vortices interaction 
energy in a multi-component BEC with no trap estimated in a different approximation (i.e., under the Abrikosov ansatz) is also proportional to $\frac{1}{r_{ij}^2}$ in the asymptotic region \cite{eto}.

Finally we note the remaining contributions, $\hat{U}_0+\hat{U}_2$, in Eq.(\ref{inter-decomp}). Their integration is quite simple and gives rise to
\begin{eqnarray}\label{U-trivial}
U_{0}&+&U_{2}\equiv \frac{g_2}{\pi^2\Delta^2}\left(\frac{N}{N_0}\right)^4( \hat{U}_0+\hat{U}_2)\nonumber\\
&\approx& \frac{g_2}{\pi\Delta}\left(\frac{N}{N_0}\right)^4\Bigg[\frac{1}{2}+2\frac{\xi^2}{\Delta}(\gamma+\ln{\frac{2\xi^2}{\Delta}})(e^{-\frac{l_i^2}{\Delta}}+e^{-\frac{l_j^2}{\Delta}})\Bigg].\nonumber\\
\end{eqnarray}
In the final expression, the first term is constant, giving no contribution to the dynamics in Eq.(\ref{lorent}), and the second one is a sum of single-particle contributions of $O(\frac{\xi^2}{\Delta})$ which renormalizes the 7-th line 
in Eq.(\ref{eq-lagrangian-final}), giving no substantial contribution to Eq.(\ref{lorent}).

%

%\newpage

%!!!!!!!!!!!!!!!!!!!!!!!!!!!!!!!!!!!!!!!!!!!!!!!!!!!!!!!!!!!!!!!!!!!!!!!!!!!!!!!!!!!!!!!!!!!!!!!!!!!!!!!!!!!!!!!!!!!!!!!!!!!!!!!!!!!!!!!!!!!!!!!!!!!!!!!!!!!!!!!!!!!!!!!!!!!!!!!!!!!!!!!!!!!!!!!!!!!!!!!!!!!!!!!!!!!!!!!!

\begin{thebibliography}{99}
\bibitem{comme} In the usual theory of hydrodynamics, interacting point vortices in 2 dimensions have 3 independent constants of motion, i.e., total energy, $x$-(or $y$-) component of momentum and $z$-component of angular momentum. Therefore the number of vortices must be larger than 3, to make the system non-integrable and chaotic.
\end{thebibliography}
\end{document}